\documentclass[10pt,journal,review]{IEEEtran}

\usepackage[pdftex]{graphicx}
\usepackage{amsmath}
\usepackage{amssymb}
\usepackage{booktabs}
\usepackage{threeparttable}
\usepackage{multirow}
\usepackage{subfig}

\begin{document}
%
\title{Deep Portrait Image \\ Completion and Extrapolation}
%
%
%
%

\author{Xian~Wu,
        Rui-Long~Li,
        Fang-Lue~Zhang,~\IEEEmembership{Member,~IEEE,}
        Jian-Cheng~Liu,
        Jue~Wang,~\IEEEmembership{Senior Member,~IEEE,}
        Ariel~Shamir,~\IEEEmembership{Member,~IEEE Computer Society,}
        and~Shi-Min~Hu,~\IEEEmembership{Senior Member,~IEEE}
\IEEEcompsocitemizethanks{\IEEEcompsocthanksitem X. Wu, R.-L. Li, J.-C. Liu and S.-M. Hu are with Tsinghua University, Beijing. S.-M. Hu is the corresponding author.
\IEEEcompsocthanksitem F.-L. Zhang is with the School of Engineering and Computer Science, Victoria University of Wellington, New Zealand.
\IEEEcompsocthanksitem J. Wang is with Megvii(Face++) Research USA.
\IEEEcompsocthanksitem A. Shamir is with the Efi Arazi school of Computer Science at the Interdisciplinary Center, Israel.}}

%
%

%

\maketitle

\begin{abstract}
General image completion and extrapolation methods often fail on portrait images where parts of the human body need to be recovered - a task that requires accurate human body structure and appearance synthesis.
We present a two-stage deep learning framework for tackling this problem.
In the first stage, given a portrait image with an incomplete human body, we extract a complete, coherent human body structure through a human parsing network, which focuses on structure recovery inside the unknown region with the help of full-body pose estimation.
In the second stage, we use an image completion network to fill the unknown region, guided by the structure map recovered in the first stage.
For realistic synthesis the completion network is trained with both perceptual loss and conditional adversarial loss. We further propose a face refinement network to improve the fidelity of the synthesized face region.
We evaluate our method on publicly-available portrait image datasets, and show that it outperforms other state-of-the-art general image completion methods. Our method enables new portrait image editing applications such as occlusion removal and portrait extrapolation. We further show that the proposed general learning framework can be applied to other types of images, e.g. animal images.
\end{abstract}


%
\IEEEpeerreviewmaketitle

\section{Introduction}

\begin{figure*}
\centering
     \subfloat[][occlusion removal]{\includegraphics[width=2.4in]{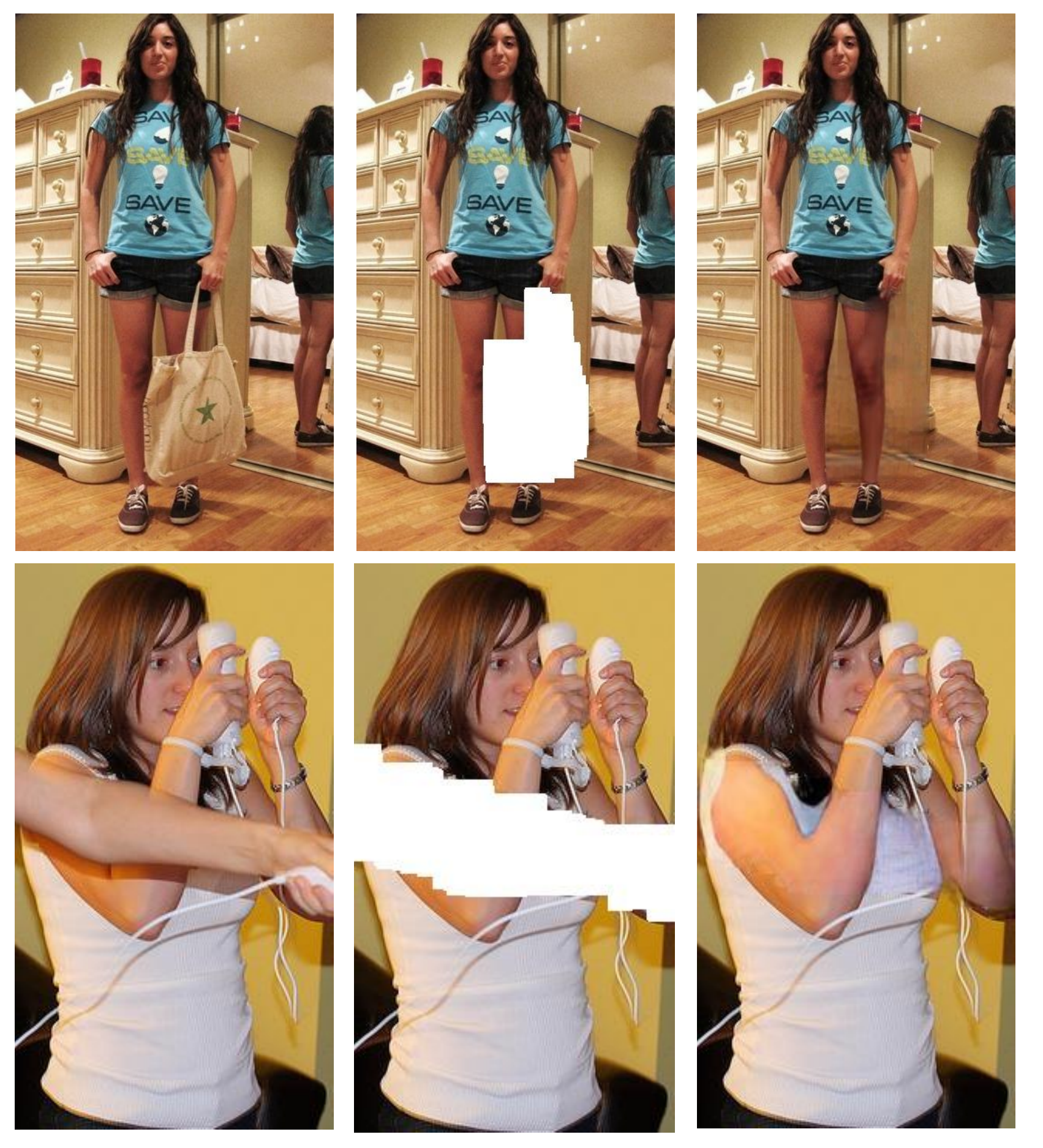}\label{fig:teaser1}}
     \subfloat[][portrait extrapolation]{\includegraphics[width=3.7in]{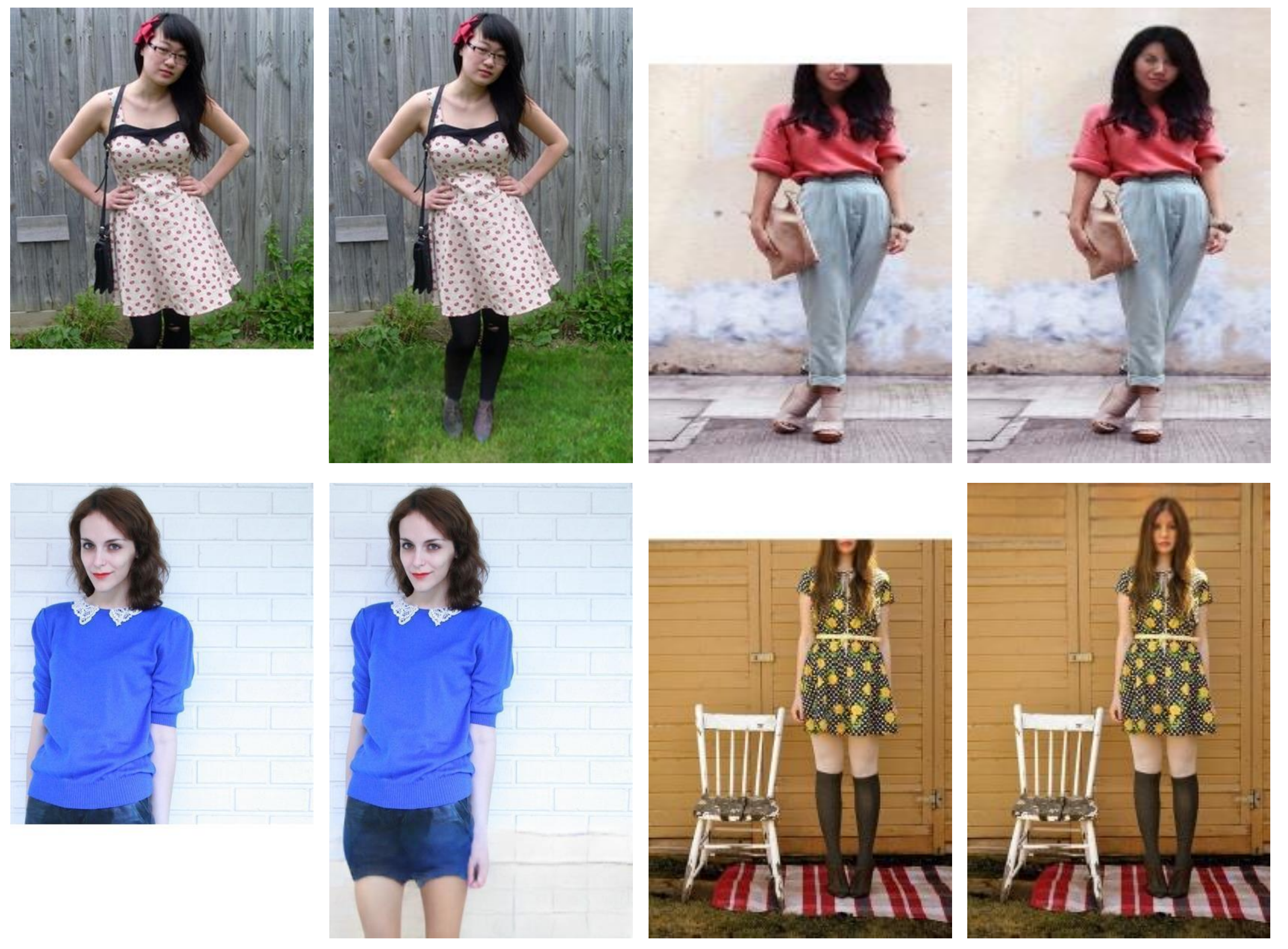}\label{fig:teaser2}}
\caption{We address the problem of portrait image completion and extrapolation. (a) shows that our method can remove the unwanted object from the portrait image. (b) shows that our method can extrapolate the portrait image to recover the lower-body or the forehead.}
\label{fig:teaser}
\end{figure*}

\IEEEPARstart{T}{here} are common mistakes that novice users often make when shooting a portrait photo.
As a typical example, while photography rules suggest that cutting off hands, feet, and foreheads can ruin the visual flow, many portrait images are taken with such improper composition. The feet are often cut off as the photographer is focusing mostly on the face region when shooting the picture (see Fig.~\ref{fig:teaser2}). At other times, the accessories that the person is carrying (e.g. the bag in Fig.~\ref{fig:teaser1}), or the other objects that partially occlude the main subject (e.g. the dog in Fig.~\ref{fig:occlusion_removal}) could be distracting and better be removed from the photo. To remove these imperfections, one could specify the unwanted objects to remove, or expand the border of the image to try to cover the whole human body, both leaving holes or blank regions in the image to be filled in properly.

Although image completion has been actively studied in the last twenty years, there is no existing approach that can work well for portrait images, where holes on the human body need to be filled in. A successful completion method is required to recover not only the correct semantic structure in the missing region, but also the accurate appearance of the missing object parts.
Both goals are challenging to achieve for portrait photos.
In terms of semantic structures, although human body has a very constrained 3D topology,
the large variation of its pose configurations, coupled with 3D to 2D projection, makes accurate structure estimation and recovery from a single 2D image difficult.
In term of appearance, despite the large variation in clothing, there are strong semantic constraints such as symmetry that the completion algorithm has to obey, e.g. two shoes usually have the same appearance regardless of how far they are separated in the image.
As we will show later, without paying special attention to these constraints, general-purpose hole filling methods often fail to generate satisfactory results on portrait photos (see Fig.~\ref{fig:ATR_completion_result} and \ref{fig:LIP_completion_result}).

Extracting human body structures from images and videos has been well studied in previous human parsing \cite{liang2015human,Gong_2017_CVPR} and pose estimation approaches \cite{newell2016stacked,cao2017realtime}. Naturally, we want to rely on these methods to estimate human body structure from an incomplete portrait photo first, and then use it to guide the image completion process.
Following this idea, we propose a two-stage deep learning framework for portrait photo completion and expansion.
In the first stage, we utilize a human parsing network to estimate both of a human pose and a parsing map simultaneously from the input image. The human pose is then used to help refine the parsing map, especially inside the unknown region.
In the second stage, we employ an image completion network to synthesize the missing region in the input image with the guidance of the parsing map,  
followed by a face refinement network for improving the generated face region. These two networks are trained sequentially with both the perceptual loss and the adversarial loss for improved realism.
We show that our approach can be used in many portrait image editing tasks that are difficult or impossible for traditional methods to achieve, such as portrait extrapolation for recovering missing human body parts (e.g. Fig.~\ref{fig:extrapolation_result}), and occlusion removal (e.g. Fig.~\ref{fig:occlusion_removal}). Furthermore, We demonstrate that the proposed learning framework is quite generic, and is applicable for other types of images such as ones of horses and cows (e.g. Fig.~\ref{fig:horse_cow_result}).

To the best of our knowledge, we are the first to integrate deep human body analysis techniques into an image completion framework. Our main contributions of this paper are summarized as follows: \begin{itemize}
\item we propose a novel two-stage deep learning framework where human body structure is explicitly recovered to guide human image completion with face refinement;
\item we show that our framework enables new portrait photo editing capabilities such as re-composition by extrapolation.
\end{itemize}

\section{Related Work}

\subsection{Image completion}

Traditional image completion methods can be categorized into diffusion-based and patch-synthesis-based ones. 
Diffusion-based methods, first proposed by Bertalmio et al. \cite{bertalmio2000image} and then extended by Ballester et al. \cite{ballester2001filling} and Bertalmio et al. \cite{bertalmio2003simultaneous}, propagate nearby image structures to fill in the holes by assuming local continuity. 
These techniques however can only handle small holes or narrow gaps. Patch-based methods are derived from texture synthesis algorithms \cite{efros1999texture,efros2001image,kwatra2003graphcut}. They extract patches from the known region of the image and use them to fill large holes. Criminisi et al. \cite{criminisi2004region} proposed a best-first algorithm by searching the most similar patch. Simakov et al. \cite{simakov2008summarizing} presented a global optimization approach based on bidirectional similarity. These techniques were greatly accelerated by PatchMatch \cite{barnes2009patchmatch,barnes2010generalized}, a randomized nearest neighbor field algorithm. 
They were further improved with appending gradients into the distance metric by Darabi et al. \cite{darabi2012image}. 
These methods can handle larger holes than propagation-based ones, but still need semantic guidance for structured scenes. 

Many inpainting approaches rely on additional guidance for semantically meaningful hole filling.
 Some use manually specified guidance, such as points of interest \cite{drori2003fragment}, lines \cite{sun2005image}, and perspective \cite{pavic2006interactive}. Other methods estimate image structures automatically, by various means such as tensor voting \cite{jia2003image}, search space constraints \cite{kopf2012quality}, statistics of patch offsets \cite{he2012statistics} and regularity in planar projection \cite{huang2014image}. However, since they only depend on low-level visual features,  such methods can only derive meaningful guidance in simple cases. 

Hays and Efros \cite{hays2007scene} presented the first data-driven method to use a large reference dataset for hole filling. 
Whyte et al. \cite{whyte2009get} extended this approach with geometric and photometric registration. Zhu et al. \cite{zhu2016faithful} presented a faithful completion method for famous landmarks using their images found on the Internet. Barnes et al. \cite{barnes2015patchtable} proposed a patch-based data structure for efficient patch query from the image database. These techniques may fail if the input image has a unique scene that cannot be found in the dataset. 

Recently, deep learning emerges as a powerful tool for image completion. However, initial approaches can only handle very small regions \cite{xie2012image,kohler2014mask,ren2015shepard}. Context encoders \cite{pathak2016context} introduced a generative adversarial loss \cite{goodfellow2014generative} into the inpainting network and combined this loss with an $L_2$ pixel-wise reconstruction loss. It can produce plausible result in an $128\times128$ image for a centered $64\times64$ hole. Yang et al. \cite{yang2017high} proposed to update the coarse result iteratively by searching nearest neural patches in the texture network for handling high resolution images. Yeh et al. \cite{yeh2017semantic} searched for the closest encoding in latent space with an adversarial loss and a weighted content loss, and then decoded it to a new complete image. More recently, Iizuka et al. \cite{iizuka2017globally} proposed an end-to-end completion network that was trained with a local and a global discriminator. The local discriminator examines the small region centered around the hole for local reality, while the global discriminator examines the entire image for global consistency. Dilated convolution layers \cite{yu2015multi} were also used to enlarge its spatial support. Yu et al. \cite{yu2018generative} extended this method with a novel contextual attention layer, which utilizes features surrounding the hole. Li et al. \cite{Li_2017_CVPR} focused on face completion with the help of semantic regularization. Existing learning-based methods are able to produce realistic completion results for general scenes such as landscapes and buildings, or certain specific objects such as human faces. However, they cannot handle portrait images as they do not consider the high-level semantic structures of human body. 

\subsection{Human parsing}

Human parsing has been extensively studied in the liturature. Liu et al. \cite{liu2015matching} combined CNN with KNN to predict matching confidence and displacements for regions into the test image. Co-CNN \cite{liang2015human} integrates multi-level context into a unified network. Chen et al. \cite{chen2016attention} introduced an attention mechanism into the segmentation network, which learns to weight the importance of features at different scales and positions. Liang et al. \cite{liang2016semantic,liang2016semantic2} employed Long Short-Term Memory (LSTM) units to exploit long range relationships in portrait images.

Without recovering the underlying human body structure, previous human parsing methods sometimes produce unreasonable results. In contrast, extracting human body structure has been extensively studied by pose estimation techniques \cite{wei2016convolutional,newell2016stacked,cao2017realtime}. Therefore, pose information has recently been integrated into human parsing frameworks to improve their performance. Gong et al. \cite{Gong_2017_CVPR} presented a self-supervised structure-sensitive learning approach. It generates joint heatmaps from the parsing result map and the ground-truth label, and then calculates the joint structure loss by their Euclidean distances. JPPNet \cite{liang2018look} builds a unified framework, which learns to predict human parsing maps and poses simultaneously. It then uses refinement networks to iteratively refine the parsing map and the pose. 
In our problem, pose estimation is helpful to guide the parsing prediction in the unknown region. Therefore, we adopt the basic architecture of JPPNet in our human parsing network, and specifically improve it for the following completion stage.

\subsection{Portrait image editing}

Previous work has explored editing portrait images in various ways, and body shape editing is one that has been paid special attention. Zhou et al. \cite{zhou2010parametric} integrated the 3D body morphable model into a single image warping approach, which is capable of reshaping a human body according to different weights or heights. PoseShop \cite{chen2013poseshop} constructed a large segmented human image database, from which new human figures can be synthesized with given poses. There are also methods for generating temporally-coherent human motion sequences with required poses \cite{xu2011video} or shapes \cite{jain2010moviereshape}. These techniques are mostly based on geometric deformation or database retrieval, so that their flexibility is limited.

Recently, many studies have been devoted to human image synthesis based on deep generative models. Lassner et al. \cite{Lassner_2017_ICCV} used the Variational Auto-Encoder (VAE) \cite{kingma2013auto} to synthesize diverse clothes in human images. FashionGAN \cite{Zhu_2017_ICCV} can change the dress of the human figure in an input image with a given text description. Zhao et al. \cite{zhao2017multi}, Ma et al. \cite{ma2017pose} and Balakrishnan et al. \cite{Balakrishnan_2018_CVPR} learned to generate human images with user-specified views or poses, while maintaining the character's identity and dress. Our goal is quite different from these methods, as we aim to recover only the missing parts of the human figure, not to generate an entirely new one.

\section{Approach}

To realistically synthesize missing human body parts, one needs to estimate plausible region-level body structure as well as coherent textures in these regions. Training a network for simultaneously predicting both the structural configuration and appearance features is extremely difficult. We instead propose a deep learning framework which employs a two-stage approach to solve this problem. In stage-I, from the incomplete human body image we predict a complete parsing map through a human parsing network. 
In stage-II, we use a completion network to generate the inpainting result with the guidance of the parsing map, and a face refinement network to further improve the face region.

\begin{figure}
\includegraphics[width=3.5in]{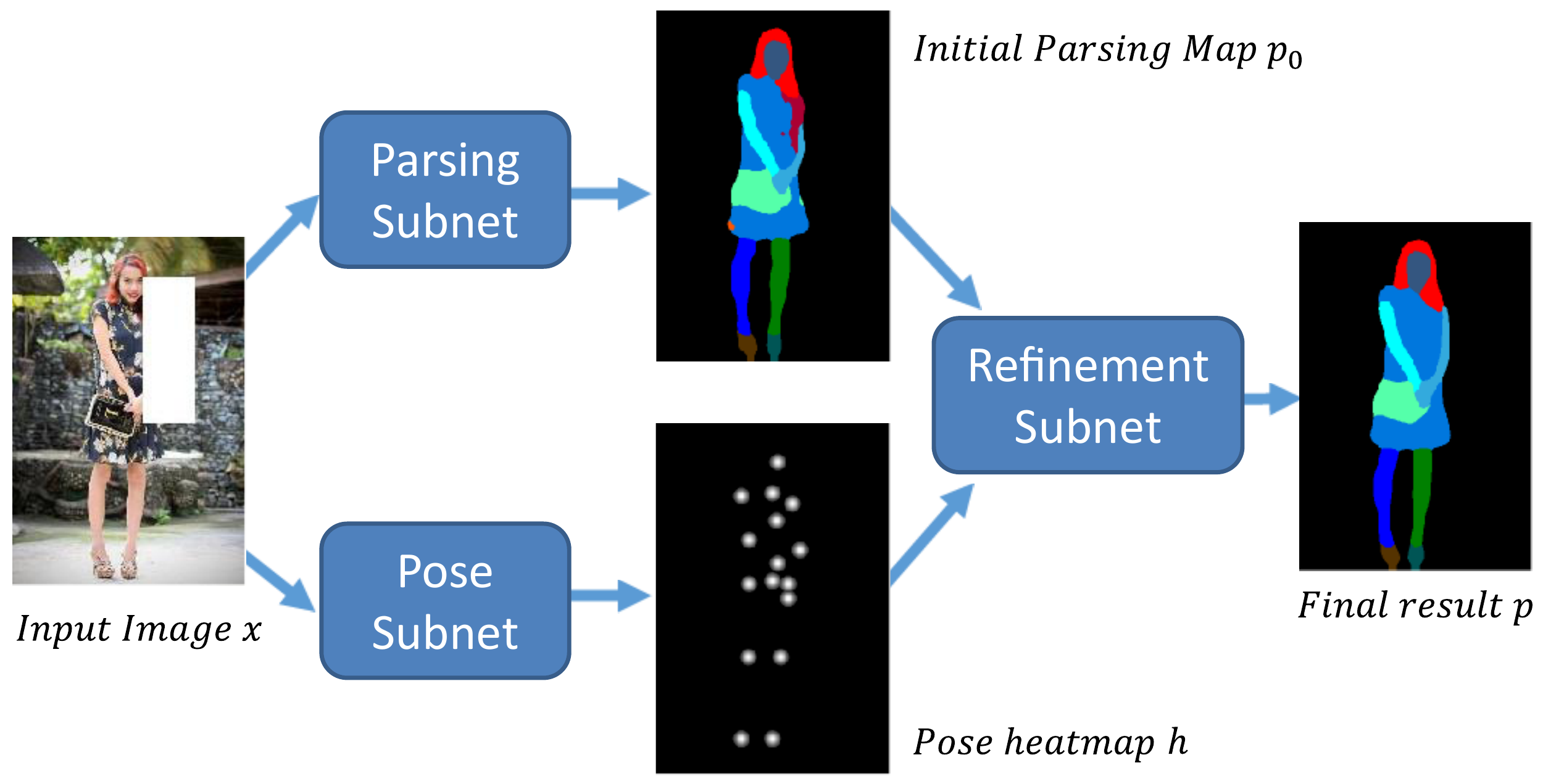}
\caption{Overview of our human parsing network at stage-I. The input image is fed into both the parsing subnet and the pose subnet. The two subnets then produce a human parsing map and pose heatmap respectively. Finally, the refinement network refines the parsing map with the help of the pose heatmap.}
\label{fig:parsing}
\end{figure}

\subsection{Stage-I: Human parsing}
Human parsing aims to segment a human body image into different parts with per-pixel labels. Compared with pose estimation, human parsing not only extracts the human body structure, but also estimates the image region boundary of each body part, which is beneficiary for image completion at the next stage. 
On the other hand, as suggested in JPPNet \cite{liang2018look}, pose estimation can help increase the accuracy of human parsing. Thus, 
we jointly train our human parsing network for both human body parsing and pose estimation. We then use the generated pose heatmap to refine the human parsing result within the unknown region. 

The input of the human parsing network is an image with a pre-defined fill-in region. We denote the input image as $\mathbf{x}$, the human parsing network as $\mathbf{P}$. Following JPPNet \cite{liang2018look}, $\mathbf{P}$ consists of two subnets, a parsing subnet and a pose subnet. The two subnets share the first four stages in ResNet-101 \cite{he2016deep}. In the parsing subnet, atrous spatial pyramid pooling (ASPP) \cite{chen2018deeplab} is applied to the fifth stage of ResNet-101 after the shared layers, to robustly segment different body parts. The parsing subnet produces the initial parsing result $\mathbf{p_0}$. In the pose subnet, several convolution layers are applied after the shared layers. The pose subnet produces the pose heatmap $\mathbf{h}$. Then a refinement subnet utilizes $\mathbf{h}$ to refine the initial parsing result $\mathbf{p_0}$ and produces the final result $\mathbf{p}$. We train the overall network end-to-end. Fig.~\ref{fig:parsing} shows the overview of our human parsing network.

For effectiveness, we make several modifications to the network architecture of JPPNet \cite{liang2018look} in $\mathbf{P}$. Specifically, we remove the pose refinement network in JPPNet and just keep the parsing refinement network. We thus do not apply iterative refinement and only refine the parsing map once. We regard $\mathbf{p}$ as our final result instead of averaging all results over different iterations. After applying these simplifications, we have found that we can generate parsing results faster with better visual quality in the unknown region.

The typical loss function of human parsing is the mean of softmax losses at each pixel, formulated as: \begin{equation} L = \frac{1}{W \times H} \sum_{i}^W \sum_{j}^H L_S(p_{ij}, \hat p_{ij}). \end{equation} $L_S$ denotes the softmax loss function. $\mathbf{\hat p}$ denotes the ground-truth labels of the human parsing output. $W$ and $H$ are the width and height of the parsing map. Importantly, in our completion task we only need to generate image content inside the unknown hole region. Thus, the parsing accuracy inside the unknown region is much more important than that of the known region. Therefore, we propose a spatial weighted loss function that gives more weights on the pixels inside the hole. It is defined as:  \begin{equation} L = \frac{1}{W \times H} \sum_{i}^W \sum_{j}^H (\alpha m_{ij} + 1)L_S(p_{ij}, \hat p_{ij}), \end{equation} where $\mathbf{m}$ is a binary mask indicating where to complete (1 for unknown pixels) and $\alpha$ is a weighting parameter.
We apply the spatial weighted loss to both the parsing subnet and the refinement subnet. We set $\alpha = 9$ in our experiments by default.

\begin{figure*}
\centering
\includegraphics[width=5.5in]{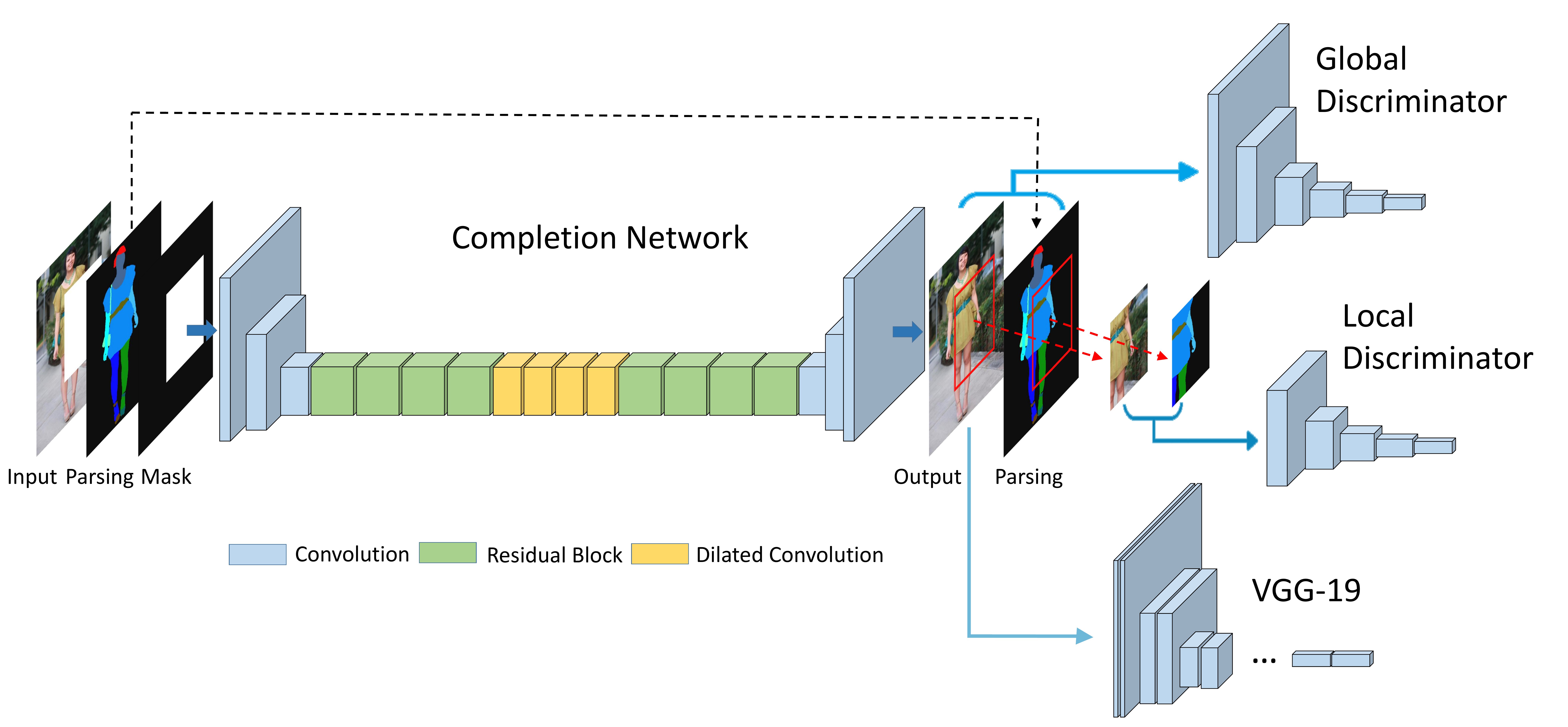}
\caption{Overview of our image completion networks at stage-II. The completion network generates the result image from the input image, parsing map and mask. It is trained with the perceptual loss and two adversarial losses. The perceptual loss measures feature maps from the VGG-19 network and the adversarial losses are backward from the discriminators. The result image with human parsing is fed into the discriminators, to tell whether it is realistic and in agreement with the structure. The input of global discriminator is the entire image with parsing, while the input of local discriminator is the local area surrounding the hole.}
\label{fig:completion}
\end{figure*}

\subsection{Stage-II: Image completion}
In stage-II, we use a completion network to synthesize missing regions with structure guidance.
The input of the completion network consists of the input image $\mathbf{x}$, the human parsing map $\mathbf{p}$ and the binary mask $\mathbf{m}$ for marking the unknown region. We denote the completion network as $\mathbf{G}$, 
and its fully convolutional architecture is shown in Fig.~\ref{fig:completion}. We bring in residual blocks \cite{he2016deep} to enhance the representation ability, and employ dilated convolutions \cite{yu2015multi} to enlarge the spatial support. Instead of using a loss based on per-pixel color distance, we use a perceptual loss measured by feature maps from a pre-trained VGG-19 network \cite{simonyan2014very}. Furthermore, to ensure that the network generates fine details that are semantically consistent with the known parts of the image, we feed the output of $\mathbf{G}$ to local and global discriminators to measure adversarial losses. The details of the architecture and our training method are described in the following sections.

\subsubsection{Completion network architecture}
The completion network begins with a stride-1 convolutional layer and then uses two stride-2 convolutional layers to downsample the resolution to $\frac{1}{4}$ of the input size.
Four residual blocks are followed to extract features of both the input image and the human parsing map. Each residual block contains two stride-1 convolutional layers with a residual connection. Residual connection \cite{he2016deep} has demonstrated its excellent ability to resolve the gradient vanishing problem, which usually happens in deep neural networks. In our experiments, we found residual blocks can significantly improve the quality of the synthesized results.
Similar to Iizuka et al. \cite{iizuka2017globally}, we use dilated convolutions \cite{yu2015multi} in the middle part of the network. Dilated convolution delivers large field of view without increasing computational cost, which is necessary for global consistency of the completed image. After that, 4 residual blocks, 2 stride-2 deconvolutional layers and a stride-1 convolutional layer are applied sequentially. Kernel sizes of the first and the last convolution layers are 7. Kernel sizes of the other convolutional layers are all 3. Batch normalization layer and ReLU activation are applied after each convolution layer except the last one. We use a tanh layer in the end to normalize the result into [-1, 1]. The completion network $\mathbf{G}$ is fully convolutional and can be adapted to arbitrary size of input images.

\subsubsection{Perceptual loss}
Most previous completion methods \cite{pathak2016context,iizuka2017globally,yu2018generative} use $L_1$ or $L_2$ per-pixel distance as loss function to force the output image to be close to the input image. Recent works \cite{chen2017photographic} have discovered that perceptual loss is more effective than $L_1$ or $L_2$ loss for synthesis tasks due to its advanced representation. Perceptual loss is first proposed by Gatys et al. \cite{gatys2016image} for image style transfer. Between a source  and a target image, it measures the distance of their feature maps extracted from a pre-trained perception network. We use VGG-19 \cite{simonyan2014very} as the perception network, which is pre-trained for ImageNet classification \cite{deng2009imagenet}.

Denote the pre-trained perception network as $\Phi$. Layers in $\Phi$ contain hierarchical features extracted from the input image. Shallower layers usually represent low-level features like colors and edges, while deeper layers represent high-level semantic features and more global statistics. Therefore, collection of layers from different levels contain rich information for image content synthesis. We define our perceptual loss function as: \begin{equation}L_p=\sum_{i=1}^n \left\|\Phi_i(\hat x)-\Phi_i(G(x,p,m))\right\|^2_2,\end{equation} where $n$ is the number of selected layers and $\Phi_i$ is the i-th selected layer. In our experiments, $\Phi$ includes \textit{relu1\_2}, \textit{relu2\_2}, \textit{relu3\_2}, \textit{relu4\_2} and \textit{relu5\_2} layers in VGG-19 \cite{simonyan2014very}.

\subsubsection{Discriminators}
Using only perceptual loss in training often leads to obvious artifacts in output images. Inspired by previous image completion methods \cite{pathak2016context,iizuka2017globally,yu2018generative}, we add an adversarial loss into the completion network to prevent generating unrealistic results. The adversarial loss is based on Generative Adversarial Networks (GANs) \cite{goodfellow2014generative}. GANs consist of two networks, a generator and a discriminator. The discriminator learns to distinguish real images from generated ones. The generator tries to produce realistic images to fool the discriminator. The two networks compete with each other, and after convergence the generator can produce realistic output.

In our task, the generated result (i.e. the filled holes) needs to be not only realistic, but also coherent with the structure guidance. Therefore, we employ conditional GANs \cite{mirza2014conditional} in our completion network, where the discriminator $\mathbf{D}$ also learns to determine whether the generated content conforms to the condition, i.e. the human parsing map $\mathbf{p}$. Hence we define the adversarial loss as: \begin{equation}L_{adv}=\min \limits_G \max \limits_D \mathbb{E}[\log D(p, \hat x) + \log (1-D(p, G(x,p,m))].\end{equation}

Inspired by Iizuka et al. \cite{iizuka2017globally}, we use both global and local discriminators in our completion network. The input to the global discriminator is a concatenation of the input image and the parsing map. The input to the local discriminator is the image region and its corresponding parsing patch centered around the hole to be completed. At training stage, the size of the image is scaled to $256\times256$ and the size of the patch is fixed to $128\times128$. As in Iizuka et al. \cite{iizuka2017globally}, we only use $\mathbf{G}$ to complete a single hole at training time. Note that $\mathbf{G}$ can fill multiple holes at once in testing stage. Noted by Pinheiro et al. \cite{pinheiro2016learning}, unequal channels of the signal may cause imbalance that one signal dominates the other ones. Therefore, we repeat the one-channel parsing map for three times to match the RGB channels. We adopt the network architecture proposed by DCGAN \cite{radford2015unsupervised} for two discriminators. All convolutional layers in the discriminators have kernels of size 4 and stride 2, except the last stride-1 convolution. The global discriminator has 7 convolution layers and the local discriminator has 6. They both end with a Sigmoid layer to produce a true or false label.
\\
\\
We define the overall loss function for the completion network as \begin{equation}L_c=\lambda_p L_p + \lambda_g L_{adv-g} + \lambda_l L_{abv-l}.\end{equation} $L_{adv-g}$ and $L_{adv-l}$ represent adversarial losses for the global and the local discriminators, respectively. $\lambda_p$, $\lambda_g$ and $\lambda_l$ are the hyper parameters to balance the different losses. We set $\lambda_p = 100$, $\lambda_g = \lambda_l = 1$ in our experiments.

\subsection{Face refinement}
The human face may only occupy a small area in the input image, but contains many delicate details that the human vision system is sensitive to. Completion network $\mathbf{G}$ can recover general missing regions well, but may have difficulties with human faces as it is not specifically trained on them. Similar to Chan et al. \cite{chan2018everybody}, we thus propose a dedicated face refinement network $\mathbf{F}$ to refine inpainted human faces.

\begin{figure}
\includegraphics[width=3.5in]{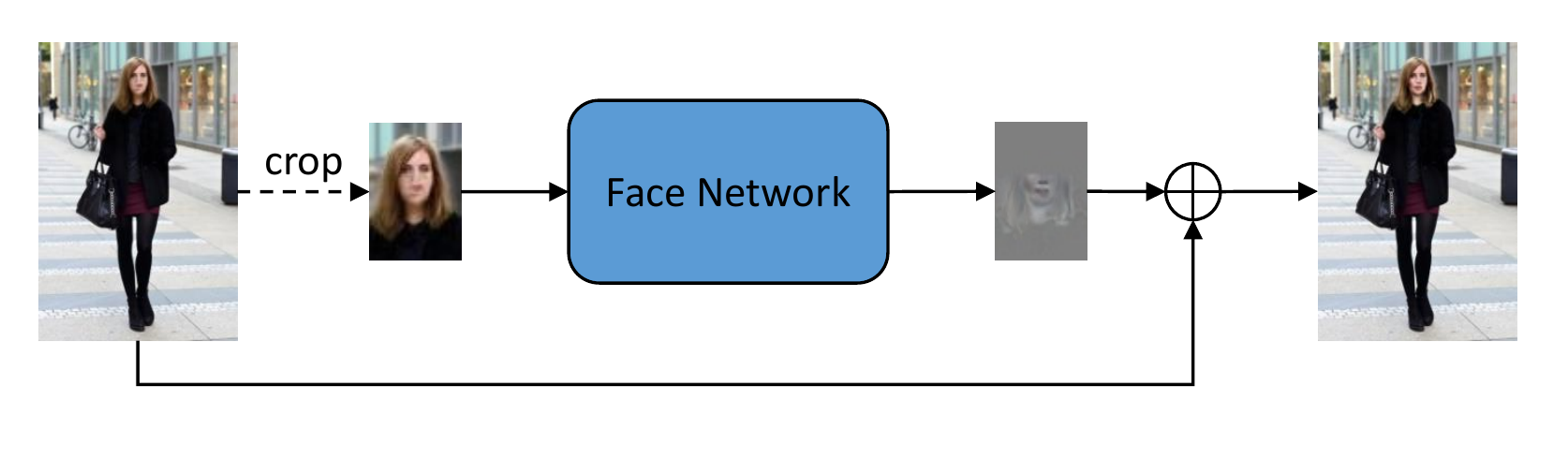}
\caption{Overview of our face refinement network. We crop the initial inpainting result of the face region and feed it into the face network. The final output is the sum of the produced residual image and the initial result.}
\label{fig:face}
\end{figure}

We crop the inpainting result $G(x,p,m)$ produced by the completion network to a window around the synthesized face $G_{f}(x,p,m)$, then feed it along with the cropped parsing map $\mathbf{p_f}$ and the cropped mask $\mathbf{m_f}$ into the face network $\mathbf{F}$. We calculate the centre of mass of the face region through the human parsing map and crop the region around the center by the size of $64\times64$. Then $\mathbf{F}$ produces the residual face image $R_{f}=F(G_{f}(x,p,m),p_{f},m_{f})$ with the size of $64\times64$. The final synthesized result $\mathbf{\hat R}$ is the addition of the residual face image $\mathbf{R_f}$ and the initial inpainting result $G(x,p,m)$. Fig.~\ref{fig:face} shows the pipeline of the face refinement.

After the training of the completion network is finished, we train the face network with the parameters of the completion network fixed. For the realism of the refined face image, we introduce a face discriminator $\mathbf{D_f}$ to distinguish generated faces from the real ones. We also use the perceptual loss to ensure that the generated faces are perceptually indistinguishable from the ground-truth. The architecture of the face network $\mathbf{F}$ is similar to the completion network $\mathbf{G}$ except that the number of the residual blocks is reduced to 4. The face discriminator $\mathbf{D_f}$ also adopts the architecture proposed by DCGAN \cite{radford2015unsupervised}. The full objective is \begin{equation}L_f=\lambda \sum_{i=1}^n \left\|\Phi_i(\hat x_f)-\Phi_i(\hat R_f)\right\|^2_2 + \log (1-D_{f}(p_f, \hat R_f)).\end{equation} We set $\lambda = 10$ in our experiments.

\subsection{Implementation details}
We train the networks at two stages separately. At stage-I, we train the human parsing network by stochastic gradient descent with momentum. We set the learning rate to 0.0001 and momentum to 0.9. At stage-II, We use the Adam solver \cite{kingma2014adam} with a batch size of 1 to train all the networks. We set the learning rate to 0.0002. We scale the input image to $256\times256$ at training time. We randomly crop a rectangular region in the input image. The edge of the rectangle is randomly set in range [64, 128]. Since we are concerned about completion for the human body, we make sure that the rectangle overlaps the human body in image. We set the pixels inside the cropped region to the mean pixel value of the datasets. To prevent overfitting, we apply several data augmentation methods for training, including scaling, cropping and left-right flipping. When training the face network, we randomly crop a rectangular region in the face area by the size of $32\times32$.

At testing time, the input image goes through the human parsing network, the completion network and the face network sequentially to generate the final result. Because all the networks are fully-convolutional, our method can be adapted to arbitrary size of image. We use Poisson image blending \cite{perez2003poisson} to post-process the final result, as previous completion methods \cite{iizuka2017globally,yang2017high} do, for more consistent boundary of the completion region.

\section{Results}
We evaluate our method on two human image datasets, ATR dataset \cite{liang2015human} and LIP dataset \cite{Gong_2017_CVPR}. ATR dataset contains 17700 human images with parsing annotations on 17 body part categories. We randomly select 1/10 of the images for testing and others for training. LIP dataset contains 30462 training images and 10000 validation images. Each image in LIP dataset has per-pixel parsing annotations on 19 categories and annotations on 16 keypoints for the pose estimation. Because there is no pose annotation in ATR dataset, we use JPPNet \cite{liang2018look} pre-trained on LIP dataset to predict the body pose for each image in ATR dataset and use the result as the ground-truth label for training.

We initialize the human parsing network with the parameters loaded from pre-trained JPPNet and we then train the network for 10 epochs on the two datasets. Meanwhile, we train the completion network for 200 epochs on ATR dataset and 100 epochs on LIP dataset. After that, we train the face network with 100 epochs on ATR dataset and 50 epochs on LIP dataset. It takes about 20 hours to train the human parsing network, 100 hours for the completion network and 50 hours for the face network. When testing on GPU, our two-stage framework spends about 0.48s to produce the final result for an input image of size $256 \times 256$. We evaluate our method on using an Intel(R) Core(TM) i7-4770 CPU @ 3.40GHz with 4 cores and NVidia GTX 1080 Ti GPU.

\begin{table}[tp]
\centering
\begin{threeparttable}
\begin{tabular}{cccccc}
\toprule
Dataset&Method&$L_1$ error&PSNR&SSIM&FID \cr
\midrule
\multirow{5}{*}{ATR}
& PatchMatch \cite{barnes2009patchmatch} & 43.0182 & 18.8905 & 0.8615 & 50.9721 \cr
& Iizuka et al. \cite{iizuka2017globally} & 12.5181 & 23.7539 & 0.9532 & 21.4838 \cr
& Yu et al. \cite{yu2018generative} & 13.7748 & 23.6954 & 0.9481 & 16.4192 \cr
& G & 10.4527 & 25.2684 & 0.9591 & 11.8323 \cr
& G+P & 9.3675 & 27.0852 & \bf{0.9736} & \bf{10.6247} \cr
& G+P+F(ours) & \bf{8.4620} & \bf{27.1511} & 0.9730 & 10.6578 \cr
\midrule
\multirow{5}{*}{LIP}
& PatchMatch \cite{barnes2009patchmatch} & 35.4193 & 20.6823 & 0.9063 & 10.9311 \cr
& Iizuka et al. \cite{iizuka2017globally} & 12.8931 & 25.0089 & 0.9646 & 9.3290 \cr
& Yu et al. \cite{yu2018generative} & 16.8708 & 22.9498 & 0.9437 & 5.3085 \cr
& G & 12.9677 & 25.1877 & 0.9630 & 3.8510 \cr
& G+P & 10.6944 & 26.0007 & \bf{0.9696} & 3.1399 \cr
& G+P+F(ours) & \bf{10.3144} & \bf{26.0641} & 0.9693 & \bf{2.9604} \cr
\bottomrule
\end{tabular}
\end{threeparttable}
\caption{Comparison with existing completion methods in terms of $L_1$ error, PSNR, SSIM and FID.}
\label{Tab:completion}
\end{table}

\begin{figure}
\centering
\includegraphics[width=3.5in]{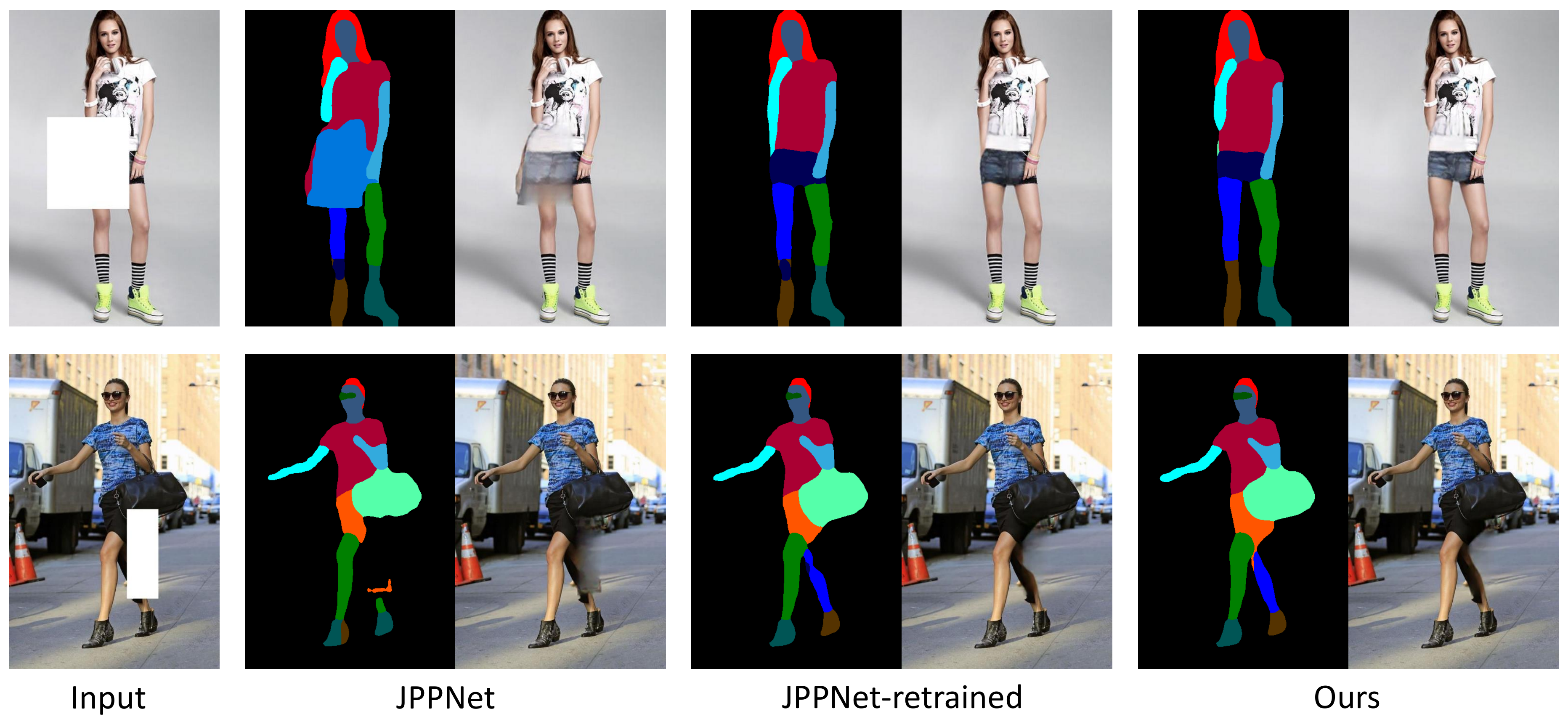}
\caption{Completion results by different human parsing methods. We complete the input image with different human parsing method. We show that our method leads to the best completion performance.}
\label{fig:parsing_comparison}
\end{figure}

\begin{figure}
\centering
\includegraphics[width=3.5in]{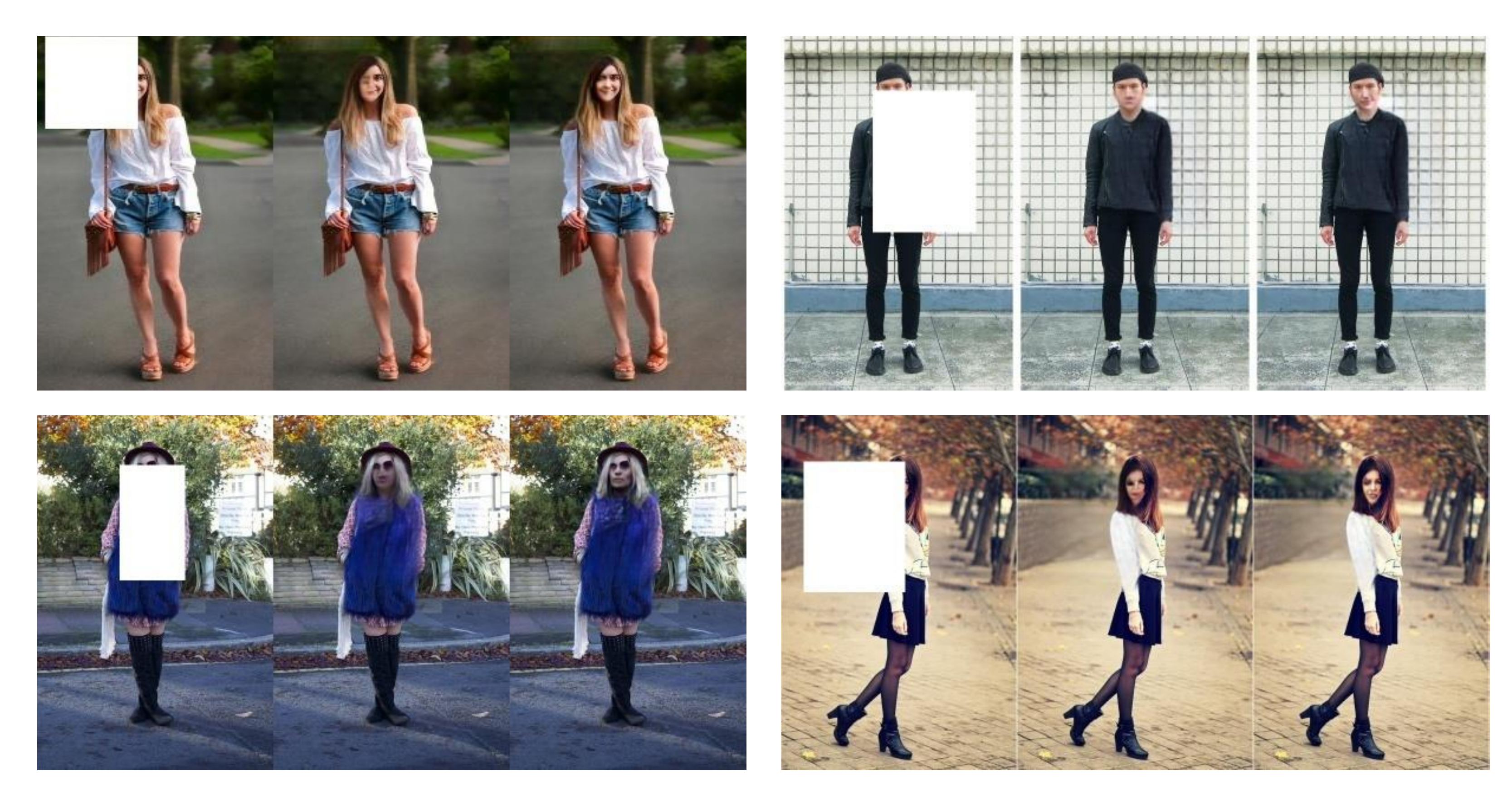}
\caption{Completion results before (middle) and after (right) the face refinement. The face network refines the facial structures and makes the completion result more realistic.}
\label{fig:face_completion}
\end{figure}

\subsection{Comparisons with existing completion methods}
\begin{figure*}
\centering
\includegraphics[width=6.5in]{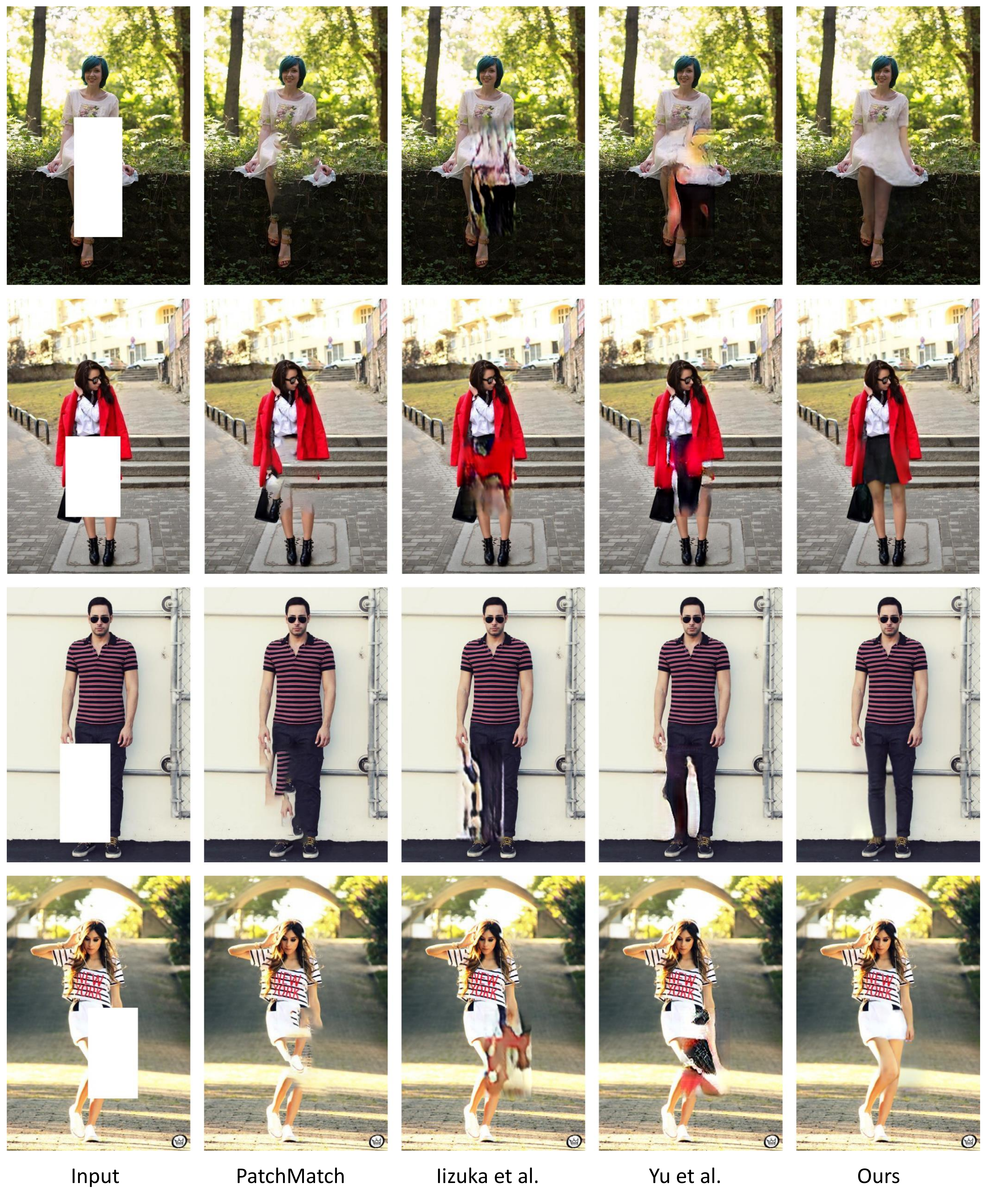}{}
\caption{Comparison results on the ATR testing set. We compare our method with PatchMatch \cite{barnes2009patchmatch}, Iizuka et al. \cite{iizuka2017globally} and Yu et al. \cite{yu2018generative}.}
\label{fig:ATR_completion_result}
\end{figure*}

\begin{figure}
\centering
\includegraphics[width=3.5in]{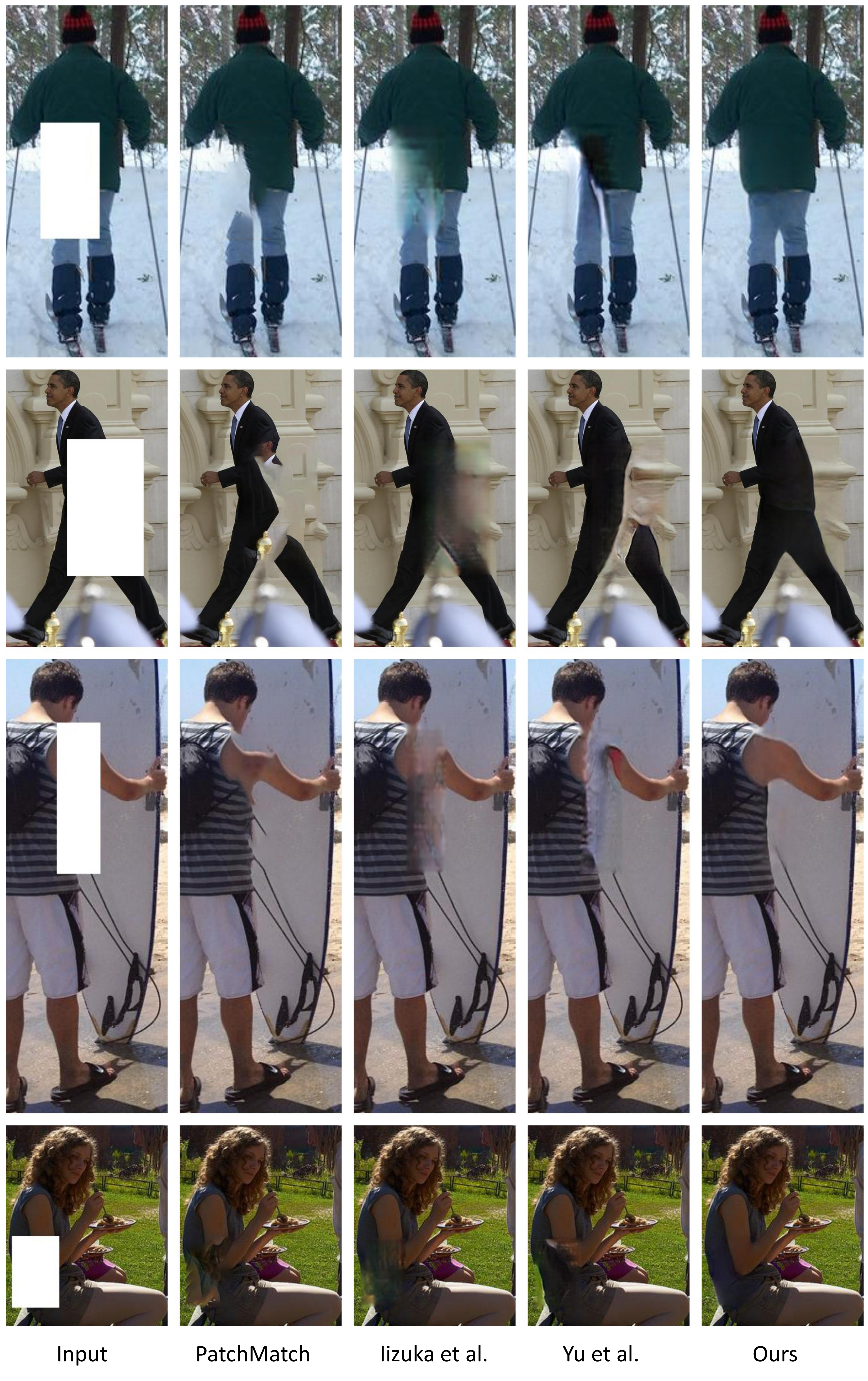}
\caption{Comparison results on the LIP validation set. We compare our method with PatchMatch \cite{barnes2009patchmatch}, Iizuka et al. \cite{iizuka2017globally} and Yu et al. \cite{yu2018generative}.}
\label{fig:LIP_completion_result}
\end{figure}

\begin{figure}
\centering
\includegraphics[width=3.5in]{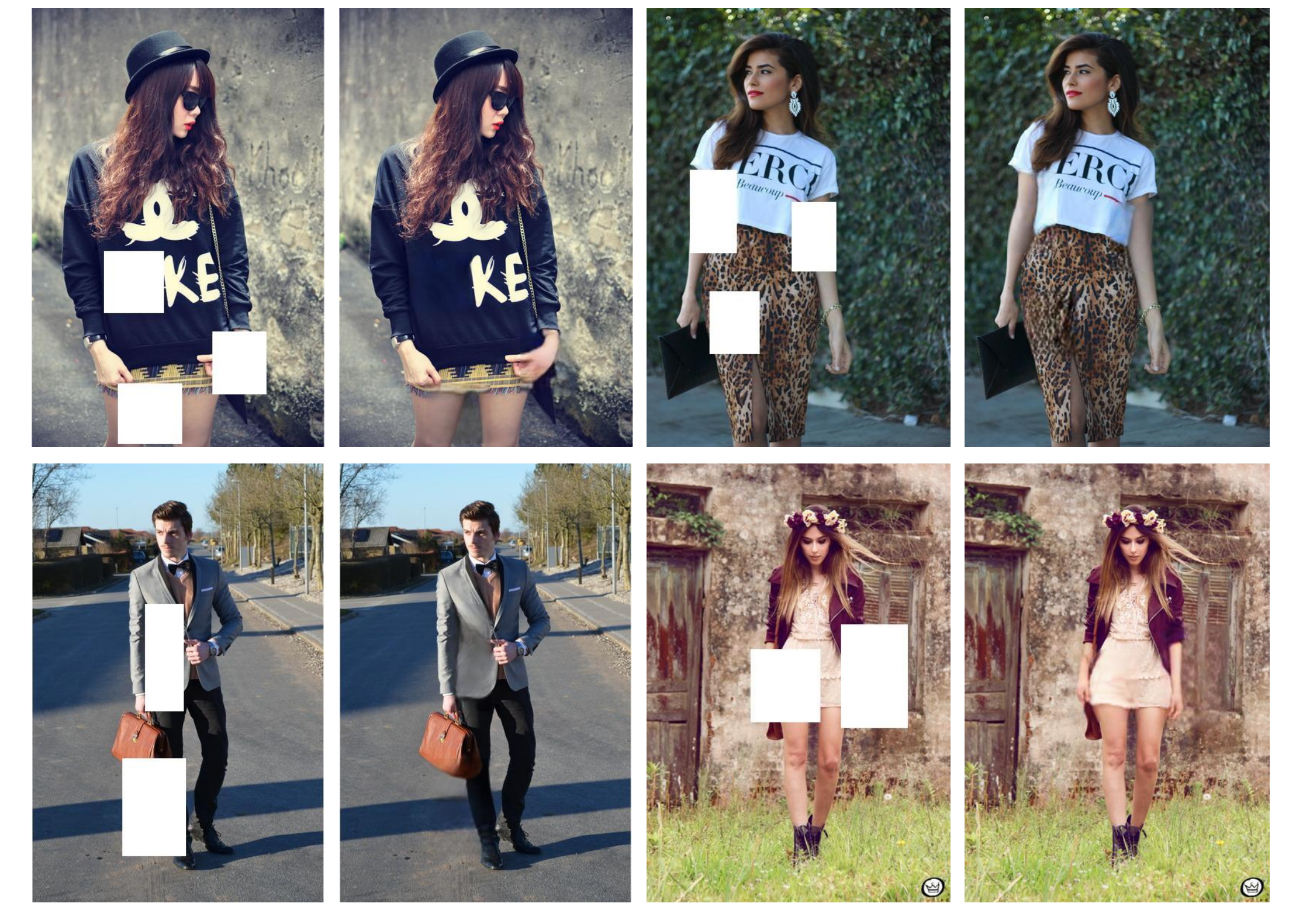}
\caption{Completion results with multiple holes on the ATR testing set.}
\label{fig:ATR_completion_m_result}
\end{figure}

We compare our approach with several existing completion methods, including Photoshop Content Aware Fill (PatchMatch) \cite{barnes2009patchmatch}, Iizuka et al. \cite{iizuka2017globally} and Yu et al. \cite{yu2018generative}. We train the model of Iizuka et al. \cite{iizuka2017globally} on training sets of ATR and LIP for the same epoch numbers as our method. We initialize the model with the parameters per-trained on Places2 dataset \cite{zhou2017places}. We also train the model of Yu et al. \cite{yu2018generative} for the same epoch and initialize it with the model pre-trained on ImageNet dataset. We test the models of Iizuka et al. \cite{iizuka2017globally}, Yu et al. \cite{yu2018generative} and our method at the scale of $256\times256$. We then resize the result to the scale of the original image for a fair comparison. Fig.~\ref{fig:ATR_completion_result} shows some comparisons from the test set in ATR. Fig.~\ref{fig:LIP_completion_result} shows some comparison results from the validation set in LIP. These results suggest that our approach can generate plausible results for large holes, while other methods fail to handle complicated human body structures. Fig.~\ref{fig:ATR_completion_m_result} also shows some completion results with multiple holes by our method.

We also compare our approach with other approaches quantitatively. We report the evaluation in terms of $L_1$ error, PSNR and SSIM \cite{wang2004image}. Table.~\ref{Tab:completion} shows that our method outperforms other methods under these measurements on the two public datasets. Please note that these evaluation metrics are by no means the best metrics for evaluating the results, since there may be many visually pleasing completion results that are equally acceptable other than the ground truth images. Recently, FID \cite{heusel2017gans} was proposed to measure the perceptual quality of the generative model. Our method also achieves the best performance under this perceptual metric.

\subsection{Comparisons on human parsing}

We compare our human parsing network with JPPNet \cite{liang2018look} to show its effectiveness for completion. We use JPPNet trained on complete images and refined on incomplete images with the same pattern as our parsing network. We apply the mean of intersection over union (IoU) as the performance metric. We measure the mean IoU for the entire image and the unknown region. We report the results in Table.~\ref{Tab:parsing}. As shown in the table, our method achieves the best parsing performance especially inside the unknown region. Fig.~\ref{fig:parsing_comparison} also shows that our human parsing method leads to the best completion result.

\begin{table}[tp]
\centering
\begin{threeparttable}
\begin{tabular}{cccc}
\toprule
Dataset & Method & Entire image & Unknown region \cr
\midrule
\multirow{3}{*}{ATR}
& JPPNet \cite{liang2018look} & 0.5193 & 0.1132 \cr
& JPPNet-retrained & 0.5786 & 0.3656 \cr
& Ours ($\alpha=0$) & \bf{0.5965} & 0.3735 \cr
& Ours & {0.5937} & \bf{0.3971} \cr
\midrule
\multirow{3}{*}{LIP}
& JPPNet \cite{liang2018look} & 0.4125 & 0.0796 \cr
& JPPNet-retrained & 0.4649 & 0.3099  \cr
& Ours ($\alpha=0$) & \bf{0.4817} & 0.3257 \cr
& Ours & 0.4685 & \bf{0.3381} \cr
\bottomrule
\end{tabular}
\end{threeparttable}
\caption{Comparison of the human parsing methods in terms of mean IoU. We measure the IoU for both the entire images and for the unknown regions on ATR dataset and LIP dataset.}
\label{Tab:parsing}
\end{table}

\subsection{Completion results for animals}

\begin{figure}
\centering
\includegraphics[width=3.2in]{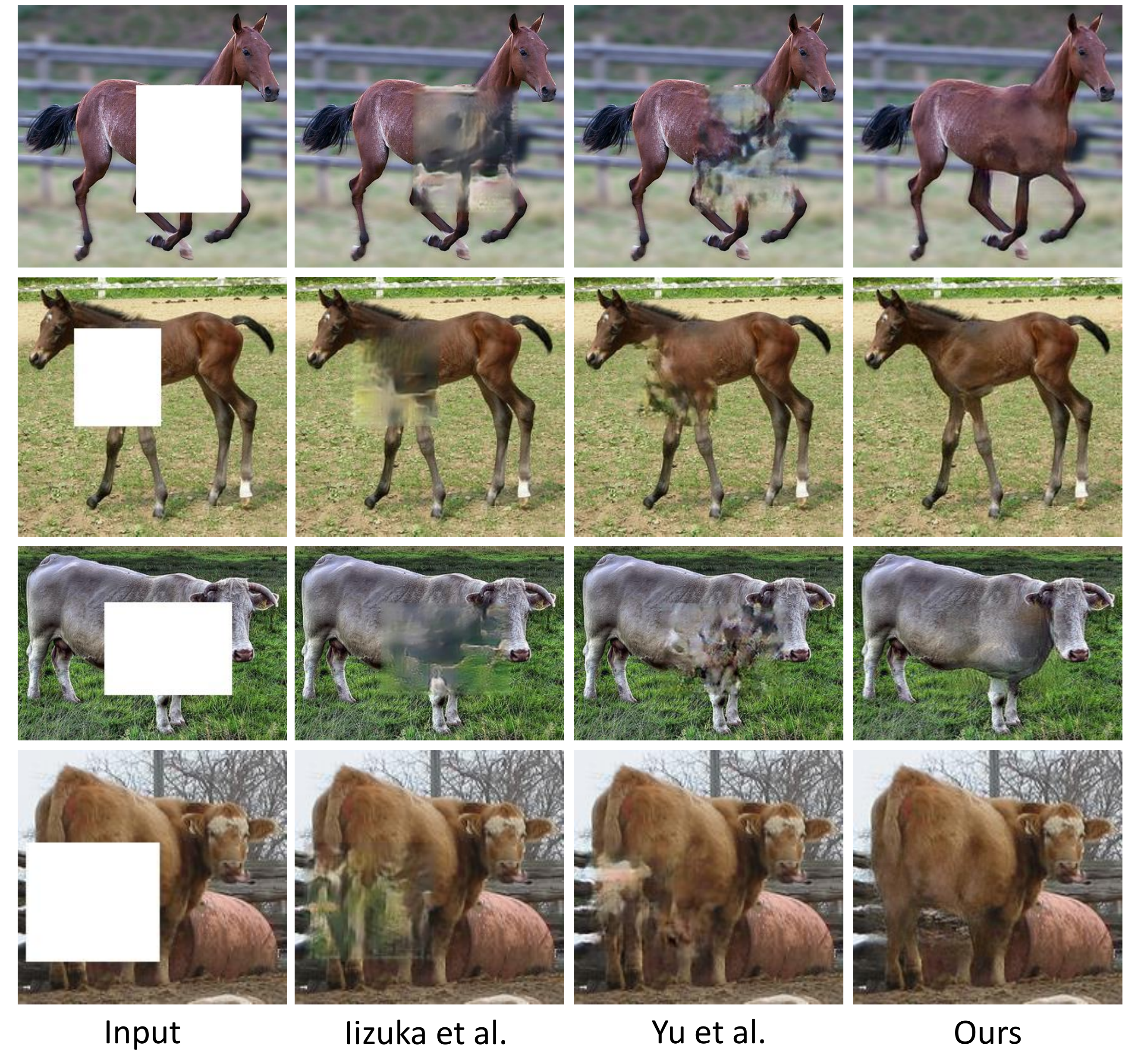}
\caption{Comparison results on Horse-Cow dataset. We compare our method with Iizuka et al. \cite{iizuka2017globally} and Yu et al. \cite{yu2018generative}. We show that our method is generic and also available for animals like horse or cow.}
\label{fig:horse_cow_result}
\end{figure}

Although we focus on human body completion in this paper, the proposed learning framework is generic and can be applied to other types of images with highly structured visual features, such as animal images. To demonstrate it, we train our model on the Horse-Cow dataset \cite{wang2015semantic}. Horse-Cow dataset contains 295 images for training and 277 images for testing. We additionally label the keypoints of poses for this dataset since it only includes parsing annotations. We re-train the parsing network for 300 epochs and the completion network for 1000 epochs on the training set. Notably, we train the horses and cows together since these two kinds of animals have similar structures. Fig.~\ref{fig:horse_cow_result} shows some completion results using the testing images. We also compare our method with Iizuka et al. \cite{iizuka2017globally} and Yu et al. \cite{yu2018generative}. We train the models of the two methods for the same epochs as ours. From the results, we can see that our two-stage deep structure can be generalized to other images in which the inherent semantic structures could be captured and predicted. 

\subsection{Ablation study}
We perform an ablation study to validate the importance of our human body structure guidance and face refinement network. We train an exactly the same completion network without using the parsing map as input. The training protocols are kept unchanged as well. This model is denoted by G. We also train a completion network with human body parsing maps, denoted as G+P. Similarly, our full model with the face refinement network is denoted as G+P+F. Results in Table.~\ref{Tab:completion} show that the parsing guidance and the face refinement are both beneficiary to the completion result. Fig.~\ref{fig:face_completion} compares the completion results before and after the face refinement, showing the effectiveness of our face network.

We also evaluate the effectiveness of our spatial weighted loss for human parsing, by comparing to the results when $\alpha$ is 0. As shown in Table.~\ref{Tab:parsing}, the spatial weighted loss can increase the parsing accuracy inside the holes, which is good for the following completion process.  

\section{Applications}

\begin{figure}[h]
\centering
\includegraphics[width=3.3in]{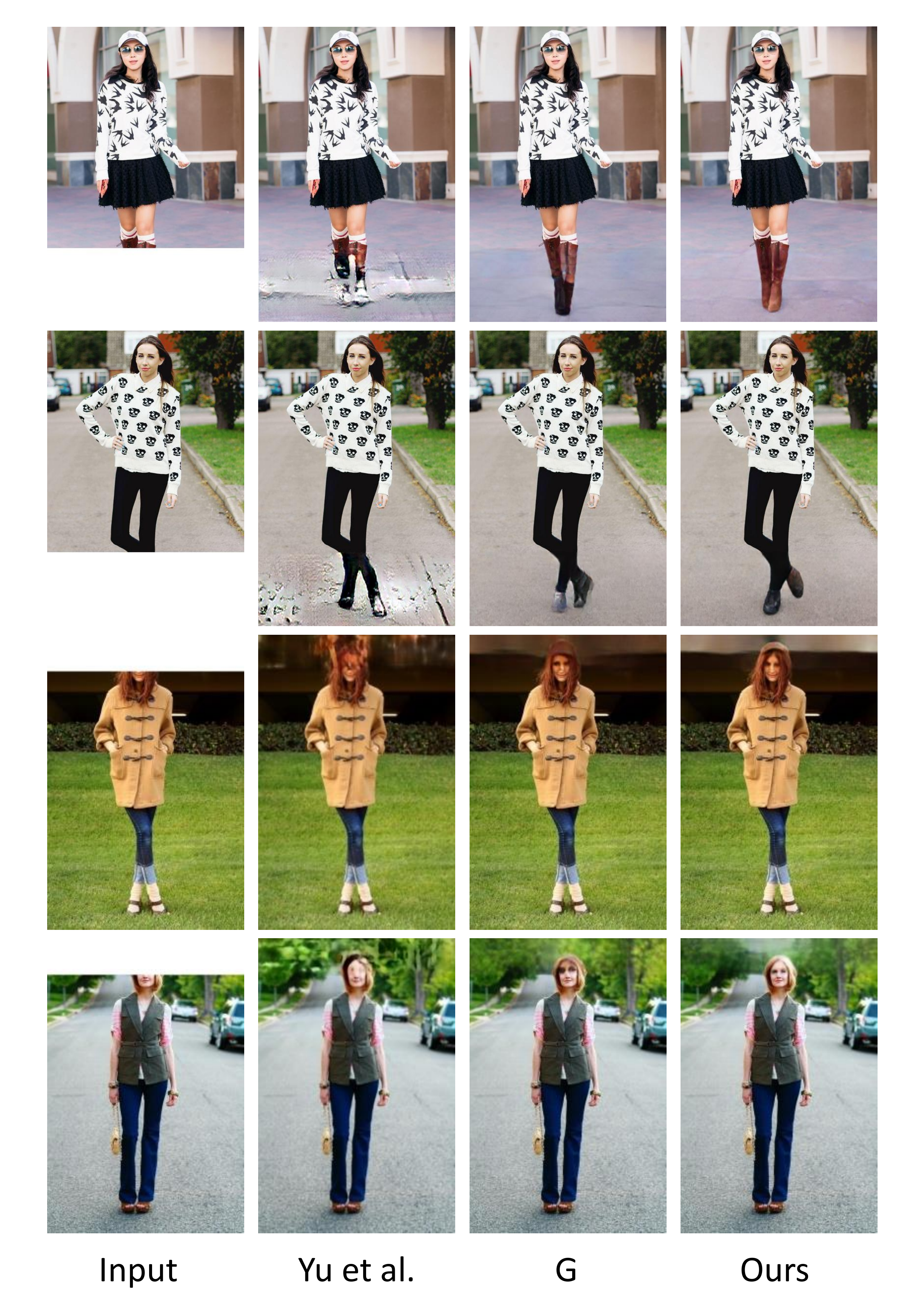}
\caption{Comparison results of portrait extrapolation. We compare our full model with Yu et al. \cite{yu2018generative} and the model G. The results show that our method can generate plausible lower body parts and faces, while other methods fail.}
\label{fig:extrapolation_result}
\end{figure}

\subsection{Portrait extrapolation}
As a general photography rule, it is not recommended to cut off people's feet or foreheads in the composition. Nevertheless, amateur photographers often make such mistakes.
It is desirable to extend the portrait image to recover the missing feet or faces to create a better composition. Fortunately, this problem can be directly solved by our framework, by treating it as a completion problem, where the unknown region is at the bottom or top part of the extrapolated image.

We re-train the entire framework by fixing the cropped rectangle in the input image. For the downward extrapolation, the bottom is cropped by the size of $64 \times 256$, and for the upward extrapolation, the top is cropped by the size of $32 \times 256$.
The reason for the different heights is that normally the missing head region is shorter than the missing legs/feet region.
We abandon the local discriminator due to the exceeded scale of the hole. We also add dropout layers into the completion network for more flexible results. Our human parsing network can analyze the posture and predict reasonable structures for the legs and shoes, as well as the hair and face. Then the completion network can generate the extrapolated image that contains consistent body parts and background. The face region would be further enhanced by the final face network. For comparisons, we also train the model of Yu et al. \cite{yu2018generative} with the same pattern. Note that, we also remove the local discriminator in their framework. To demonstrate the importance of the structural information in the portrait extrapolation task, we train the model G without the parsing guidance as a comparison, as shown in Fig.~\ref{fig:extrapolation_result}. These results suggest that by generating high fidelity human body completion results, our method opens new possibilities for portrait image enhancement and editing.

\subsection{Occlusion removal}
\begin{figure}
\centering
\includegraphics[width=2.6in]{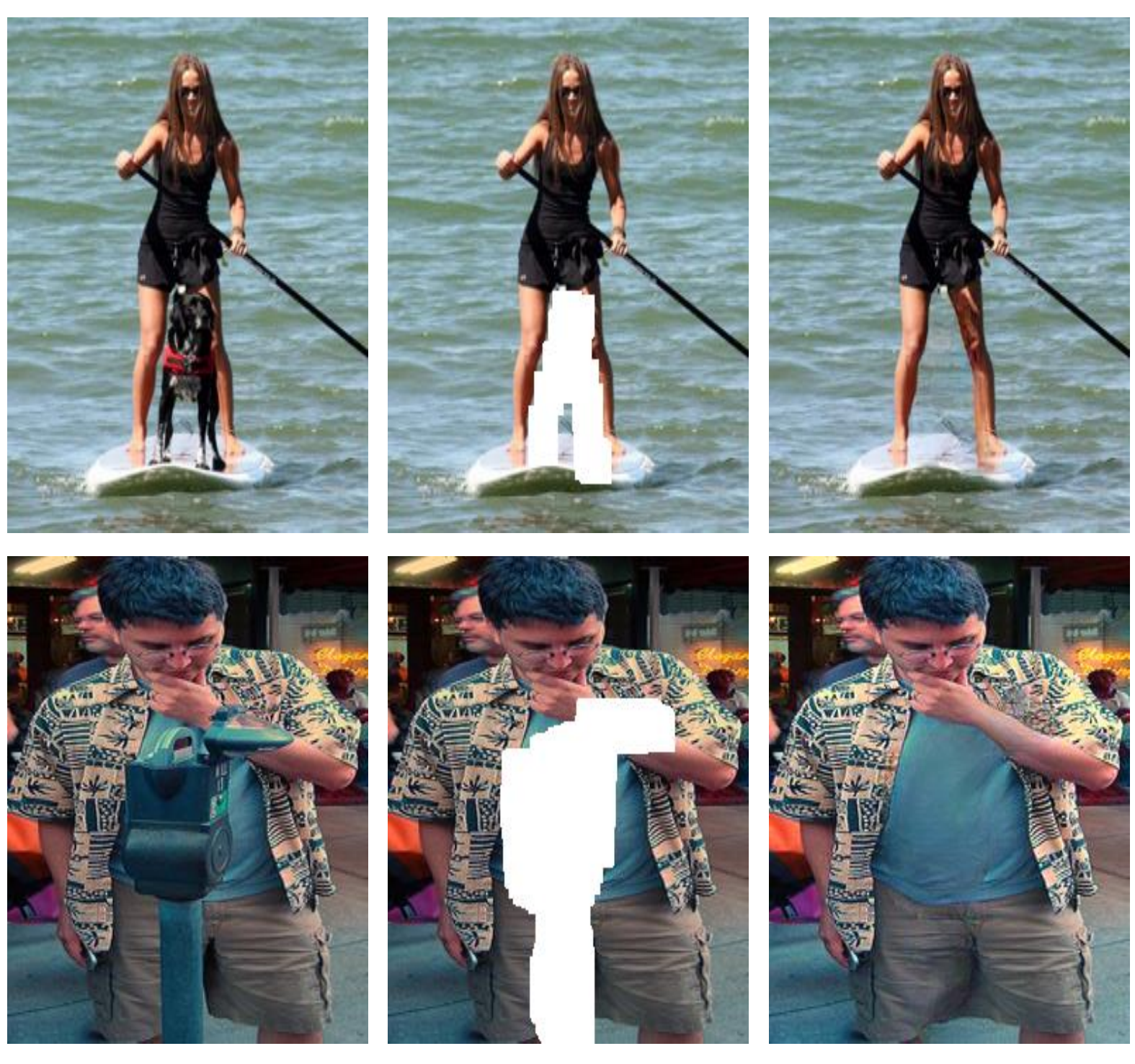}
\caption{Examples of occlusion removal by our approach.}
\label{fig:occlusion_removal}
\end{figure}

Occlusion removal is a natural application for the task of image completion. Fig.~\ref{fig:occlusion_removal} shows that our approach can recover the full human body when removing the unwanted objects. Because our framework has a human parsing stage to generate correct structural information for the hole region, we can handle large occlusions as shown in the bottom example in Fig.~\ref{fig:occlusion_removal}. Note that we preserve the geometric features along the arm and clothes boundaries after completion.

\section{Conclusion and Limitations}

We propose a two-stage deep learning framework to solve portrait image completion problem. We first employ a human parsing network to extract structural information from the input image. 
Then we employ a completion network to generate the unknown region with the guidance of the parsing result and a following face network to refine the face appearance.
We have demonstrated that, aware of the structure of the human body, we can produce more reasonable and more realistic result compared to other methods. And we have shown the capability of our method for applications like occlusion removal and portrait extrapolation. Besides humans, we have also experimented our method on animals and achieved impressive completion results, which indicates that our framework can be extended to other kinds of images where the inherent semantic structures could be encoded and predicted.

\begin{figure}
\centering
\includegraphics[width=3.2in]{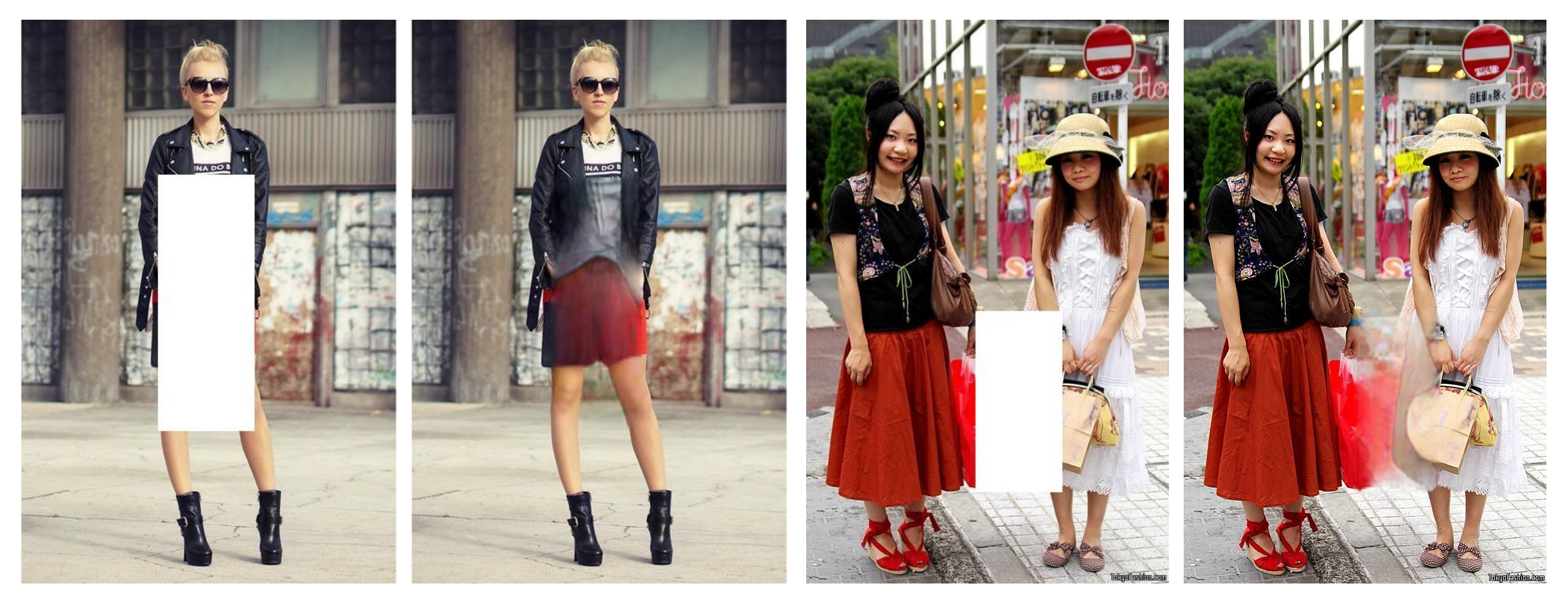}
\caption{Some failure cases from the ATR dataset. Our method cannot produce satisfactory result when the textured region is heavily covered and will be confused with multiple persons.}
\label{fig:failure}
\end{figure}

Our method may fail in some cases as shown in Fig.~\ref{fig:failure}.
Firstly, when most of the textured region or the logo is covered, our method may not yield satisfactory completion result due to the inadequate information, as shown in Fig.~\ref{fig:failure} (left).
Secondly, our model will be confused when the hole region is connected to multiple persons, as shown in Fig.~\ref{fig:failure} (right). That is because our current human parsing network is only trained by portrait images with a single person. We plan to extend our framework to deal with multiple persons in the future.

\section*{Acknowledgment}

The authors would like to thank all the reviewers. This work was supported by National Natural Science Foundation of China (Project Number 61561146393 and 61521002). Fang-Lue Zhang was supported by Research Establishment Grant of Victoria University of Wellington (Project No. 8-1620-216786-3744). Ariel Shamir was supported by the Israel Science Foundation (Project Number 2216/15).

\ifCLASSOPTIONcaptionsoff
  \newpage
\fi



%

\bibliographystyle{IEEEtran}
\bibliography{template}

\end{document}